\documentclass[12pt]{article}
\pdfoutput=1

\usepackage{epsfig}
\usepackage{graphicx}
\usepackage{color}

\usepackage[numbers,compress]{natbib}
\usepackage[T1]{fontenc} 
\usepackage{amsmath,amssymb,mathrsfs}
\usepackage{subfigure}
\usepackage[colorlinks=true
,urlcolor=blue
,anchorcolor=bleu
,citecolor=blue
,filecolor=blue
,linkcolor=blue
,menucolor=blue
,linktocpage=true
,pdfproducer=medialab
,pdfa=true
]{hyperref}

\def\be{\begin{eqnarray}}
\def\ee{\end{eqnarray}}
\def\beq#1\eeq{\begin{align}#1\end{align}}

\def\la{\label}
\def\eq#1{\eqref{#1}}

\def\b{\beta}

\def\d{\delta}
\def\D{\Delta}
\def\e{\epsilon}

\def\r{\rho}

\def\pa{\partial}
\def\fr{\frac}

\def\lb{\left[}

\def\rb{\right]}

\newcommand{\nn}{ \\ \nonumber}
\newcommand{\al}[1]{\begin{align}#1\end{align}}

\begin{document} 
\flushbottom
\begin{titlepage}

\hfill\parbox{5cm} { }

\vspace{25mm}

\begin{center}
{\Large \bf Dual Geometry of Entanglement Entropy \\ via Deep Learning}

\vskip 1. cm
Chanyong Park$^a$\footnote{e-mail : cyong21@gist.ac.kr},
Chi-Ok Hwang$^b$\footnote{e-mail : chwang@gist.ac.kr},
Kyungchan Cho$^a$\footnote{e-mail : kcho9803@gist.ac.kr},
and Se-Jin Kim $^a$\footnote{e-mail : sejin-\,\!-\,\!-@gist.ac.kr}

\vskip 0.5cm

{\it  $^a\,$Department of Physics and Photon Science, Gwangju Institute of Science and Technology, Gwangju 61005, Korea}\\
{\it $^b\,$Division of Liberal Arts and Sciences, GIST College, Gwangju Institute of Science and Technology, Gwangju 61005, South Korea}\\
\end{center}

\thispagestyle{empty}

\vskip1cm


\centerline{\bf ABSTRACT} \vskip 4mm

For a  given entanglement entropy of QFT, we investigate how to reconstruct its dual geometry by applying the Ryu-Takayanagi formula and the deep learning method. In the holographic setup, the radial direction of the dual geometry is identified with the energy scale of the dual QFT. Therefore, the holographic dual geometry can describe how physical properties of a quantum field theory change along the RG flow. Intriguingly, we show that the reconstructed geometry only from the entanglement entropy data can give us more information about other physical properties like thermodynamic quantities in the IR region.

\vspace{2cm}


\end{titlepage}


\section{Introduction}

Recently, the AdS/CFT correspondence \cite{Maldacena:1997re,Gubser:1998bc,Witten:1998qj,Witten:1998zw,Aharony:1999ti}, which maps a $d$-dimensional conformal field theory (CFT) to a $(d+1)$-dimensional anti-de Sitter (AdS) space, has been widely investigated to understand the strongly coupled quantum field theory (QFT). 
Furthermore, its generalization called the gauge/gravity duality has been applied to the study of renormalization group (RG) flow by deforming a CFT \cite{Henningson:1998gx,Henningson:1998ey,Freedman:1998tz,Gubser:1999vj,deBoer:1999tgo,Skenderis:1999mm,Heemskerk:2010hk,Balasubramanian:1999re,Park:2021nyc}. In this holographic study, a local operator deforming the field theory is realized by a bulk field modifying the background AdS space. For the ABJM theory \cite{Gustavsson:2007vu,Aharony:2008ug}, for instance, a mass deformation makes a UV CFT change into another IR CFT along the RG flow. The authors of Ref. \cite{Bobev:2018wbt} figured out the holographic RG flow connecting two fixed points by using the solution of the BPS equations. 

The authors of Ref. \cite{Hashimoto:2018ftp} constructed the field equation of an AdS space as a neural network (NN) and showed the duality between the deep learning (DL) and the AdS space. For the conventional DL, the deep layers are usually considered as a black box we cannot understand. However, knowing the black box is important to understand the underlying structure of the system. Reconstructing the dual gravity following the AdS/CFT correspondence may give us a hint to understand the black box. In Ref. \cite{Hashimoto:2018ftp,Song:2020agw,Dolotin:2022cev,Bao:2022rup,Hirst:2022qqr}, the authors utilized deep layers satisfying a specific recursion relation and determines the dual bulk geometry.

The holographic or AdS/CFT correspondence claims that a $(d+1)$-dimensional classical gravity is dual to a $d$-dimensional QFT. In this case, the radial or extra dimension of the gravity is identified with the RG flow of the dual QFT. To construct the gravity from the QFT data, therefore, the QFT data must contain the information about the energy scale dependence. The entanglement entropy is one of the important quantities representing quantum nature of QFTs like quantum correlation. Another important feature of the entanglement entropy is that it can describe the real space RG flow of QFTs. Therefore, the entanglement entropy is useful to reconstruct the dual gravity from the QFT's data \cite{Swingle:2014uza,Lashkari:2013koa,Faulkner:2013ica,Nozaki:2013vta}. In the holographic study, the entanglement entropy is realized by a minimal surface extending from the boundary to the bulk \cite{Ryu:2006ef,Ryu:2006bv,Nozaki:2012zj,Casini:2008cr,Blanco:2013joa,Nishioka:2009un}.

The minimal surface defined in a three-dimensional black hole-type geometry describes the RG flow of a thermal two-dimensional QFT. In this case, the three-dimensional bulk metric has the following general form
\beq
ds^2 = \frac{R^2}{u^2} \left( -f(u)dt^2 + \frac{du^2}{f(u)} + dx^2\right).
\eeq
The entanglement entropy is given by the area of the minimal surface living in this geometry. For a three-dimensional geometry, the area of the minimal surface reduces to a geodesic length. Even in this case, it is hard to predict the details of the geometry, $f(u)$, from the known entanglement entropy data. This comes from the fact that   the entanglement entropy is usually given by an integral form of the bulk metric. Thus, we can not directly figure out the geometry from the entropy function even if the entanglement entropy has a simple form. In general, perturbation is very useful to analyze complicated mathematical structures in some regions. When the geometry is deformed by a relevant operator \cite{Kim:2019feb,Kim:2019ewv,Kim:2019rwd,Kim:2020unz}, we can compute the entanglement entropy perturbatively in the asymptotic AdS space which corresponds to the conformal perturbation theory on the dual QFT side. Inversely, reconstructing the perturbative geometry from entanglement entropy data is also possible. However, this perturbative reconstruction is valid only in the UV regime. To know the dual geometry in the entire range, we must go beyond the perturbative reconstruction.

The goal of this work is to find nonperturbatively a  function $f$ reproducing the known entanglement entropy data. To do so, we exploit the DL method~\cite{Hashimoto:2018ftp,Song:2020agw}. After introducing the deep layers, we make the general recurrence relation of the subsystem size $l$ and the entanglement entropy $S_E$. 
By optimizing the function $f$ after defining a loss function appropriately, we can finally reconstruct the nonperturbative dual geometry from given entanglement entropy data. We show that the holographic entanglement entropy in this nonperturbative geometry reproduces the original entanglement entropy data, as it should do. Knowing the dual geometry is equivalent to knowing the underlying theoretical structure of the dual QFT, as mentioned before. Intriguingly, this underlying structure allows us to get more information of the system. For example, if thermalization scale is much higher than other scales of a system,  
the entanglement entropy is reduced to the thermal entropy in the IR region. In this case, the reconstructed dual geometry can determine all other thermodynamic quantities, like temperature, internal energy, and pressure. These quantities are nonperturbative results appearing in IR region of the RG flow.


\section{Thermodynamics of Schwarzschild-type black holes}

Holographic principle is one of the fascinating tools to understand strongly interacting systems. Recently, there have been many attempts to figure out various nonperturbative features of QFT in a gravity theory of one higher dimension. Unfortunately, the exact holographic relation was known
only for maximally supersymmetric and conformal field theories, like $N=4$ super Yang-Mills and ABJM theories.  To overcome this limitation, it would be important to clarify dual gravity theories of nonconformal systems. In the present work, we study how to reconstruct the dual geometry of QFT from the entanglement entropy data. The reconstructed dual geometry allows us to understand other physical properties, as we will see later.

Before studying the reconstruction of the dual geometry, let's first discuss how one can relate the dual geometry to the entanglement entropy in the holographic setup. We first assume a two-dimensional thermal system which has no other scale except temperature. Then, its holographic dual can be described by the following three-dimensional metric 
\be
ds^2 = \frac{R^2}{u^2} \left( - f(u) dt^2 +\frac{du^2}{f(u)} + d x^2 \right) .   \la{Metric:BH}
\ee
This is one of the metric ansatz representing an asymptotic AdS space whose dual two-dimensional QFT has a UV fixed point. To have an asymptotic AdS space, the unknown metric function $f(u)$ should be one at $u = 0$. For a pure AdS space, the metric factor $f(u)$ is given by $f(u) = 1$.

Another example allowing the same metric ansatz is a black hole in the Einstein frame. For the Schwarzschild black hole, a blackening factor is given by
\be
f (u) = 1  -   \fr{u^2}{u_h^2}   .
\ee
The blackening factor allows a simple root $u_h$ called the horizon. In the outside of a black hole ($0 \le u  < u_h$), the blackening factor is always positive. Intriguingly, it was known that the quantities characterizing a black hole satisfy the thermodynamics law. From the holography point of view, the black hole thermodynamics corresponds to that of the dual QFT. It was also known that a $p$-brane gas uniformly distributed in an AdS space admits a Schwarzschild-type black hole with the following blackening factor
\be
f(u) = 1 - \fr{u^{2-p}}{u_h^{2-p}}  , \label{pbrane f}
\ee
where the horizon $u_h$ crucially relies on the energy density of a $p$-brane gas \cite{Park:2021wep,Park:2021tpz}.

Due to the existence of a horizon for a Schwarzschild-type black hole solution, a blackening factor can be reexpressed as
\beq
f(u) = \left( 1- \frac{u}{u_h}\right)  \  g (u ),   \la{Ansatz:SBhole}
\eeq
where $g(u)$ is a function of a dimensionless variable, $u/u_h$, and always positive outside the horizon. When a black hole is characterized by only one parameter like a black hole mass. the blackening factor of a Schwarzschild-type black hole has a fixed value at the horizon which is independent of the horizon's position. In this case, all thermodynamic quantities of the black hole is determined by the Hawking temperature and Bekenstein-Hawking (or thermal) entropy. For a Schwarzschild-type black hole, the Hawking temperature $T$ and Bekenstein-Hawking  entropy $S$ are given by
\beq
T & =\frac{g(u_h)}{4 \pi u_h}, \\
S & = \frac{R \, L }{4G u_h } , \la{Result:thentropy}
\eeq
where $L$ corresponds to an appropriately regularized one-dimensional volume. The temperature and entropy together with the thermodynamics law determine an internal energy of the thermal system \cite{Saha:2019ado,Kim:2016jwu}
\beq
E &= \int T \ dS =\frac{d-1}{d} T S .
\label{energy}
\eeq

These thermodynamic quantities can further fix other physical properties. From now on, we focus on the $d=2$ case for convenience. Then, the above thermodynamic quantities determine a free energy and pressure as the following form
\beq
F &= E - T S= -\frac{R L }{32 \pi G } \frac{g(u_h)}{u_h^2},
\nn\\
P &= - \fr{\pa F}{\pa L} = \fr{R }{32 \pi G } \frac{g(u_h)}{u_h^2} , 
\eeq
where $V_1 = L$ corresponds to the system size. Then, the equation of state parameter of this system reads
\beq
w &= L \fr{\pa P}{\pa E} =1.
\eeq
This corresponds to that of massless field or radiation for a two-dimensional QFT. Lastly, a heat capacity becomes
\beq
c_V & = \fr{R L}{4 G } \frac{1}{u_h} >0 . 
\eeq
The positivity of the heat capacity indicates that the thermal system considered here is thermodynamically stable.

If we take into account a black hole with more hairs like charged or rotating black holes, the value of $g(u)$ at the horizon usually depends on the hairs. To determine thermodynamics of this system, we need to know further the parameter dependence of $g(u)$. Hereafter, we concentrate on a Schwarzschild-type black for simplicity, though the technique studied in this work is also applied to a black holes with multiple hairs.

\section{Thermodynamics from the Entanglement entropy}

In the previous section, we discussed how to understand various thermodynamic properties from  black hole geometries. Such thermodynamic quantities are also understood from the quantum entanglement entropy. Since the entanglement entropy explains a real space RG flow, the thermal entropy discussed before appears as IR physics of the entanglement entropy \cite{Kim:2016jwu,Kim:2016hig,Park:2015hcz}. In general, the entanglement entropy suffers from UV divergences. After removing the UV divergences with an appropriate renormalization scheme, the renormalized entanglement entropy satisfies the area law in the UV region. For the dual QFT of a black hole, however, the renormalized entanglement entropy in the IR region shows the volume law. This is because the leading contribution to the entanglement entropy in the IR regime comes from the thermal entropy following the volume law \cite{Kim:2016jwu}. As a result, the IR behavior of the entanglement entropy gives us information about the thermal entropy. When a black hole geometry is known, one can easily calculate the entanglement entropy following the Ryu-Takayanagi (RT) proposal. According to the AdS/CFT correspondence, the entanglement entropy and its dual geometry must have a one-to-one correspondence \cite{Swingle:2014uza,Lashkari:2013koa,Faulkner:2013ica,Nozaki:2013vta,Ryu:2006ef,Ryu:2006bv,Nozaki:2012zj,Casini:2008cr,Blanco:2013joa,Nishioka:2009un}. Therefore, it must be possible to reconstruct a dual geometry from the given entanglement entropy \color{red}\cite{Bilson_2011,Spillane:2013mca,Jokela_2021,Bao_2020}\color{black}. In this section, we first discuss how to evaluate the entanglement entropy of the given geometry. By considering the inverse procedure of the RT formula in the next sections, we will investigate how to reconstruct the dual geometry of the given entanglement entropy.

To calculate the holographic entanglement entropy, we consider the following three-dimensional asymptotic AdS space
\beq
d s^2 = \frac{R^2}{u^2} \left( -f(u) d t^2 +\frac{d u^2}{f(u)} + d x^2\right). \label{metric}
\eeq
and divide its boundary into two parts, a subsystem and its complement. Parameterizing the subsystem size as $- l /2 \le x \le l /2$ at $u=0$, the entanglement entropy is given by the area of a minimal surface extending to the dual geometry. Following this conjecture, the metric in \eq{Metric:BH} yields the following holographic entanglement entropy
\be
S_E = \fr{1}{4 G} \int_{-l/2}^{l/2} dx \  \frac{R}{u} \fr{\sqrt{f (u) + u'^2}}{ f (u)} ,
\ee
where the prime means a derivative with respect to $x$. Here, we focus on a connected minimal surface to describe the entanglement entropy depending on the subsystem size. In this case, the translation symmetry in the $x$-direction gives rise to a conserved quantity
\al{
H = -\frac{R}{u} \frac{\sqrt{f}}{\sqrt{f + u'^2}} .
}
Recalling that the entanglement entropy is invariant under $x \to -x$, the minimal surface should have a turning point at $x=0$. Denoting the turning point as $u_0$, $u'$ becomes zero at $u=u_0$ and the turning point provides a maximum value to which the minimal surface can extend. In other words, the minimal surface extends to only the range of $0 \le u \le u_0$. At the turning point, the conserved quantity reduces to
\al{
H = -\frac{R}{u_0}.
}
Using this relation, we can represent the subsystem size and the entanglement entropy in terms of the turning point 
\beq
l & = \int_0^{u_0} du  \ \fr{2u}{\sqrt{f} \, \sqrt{u_0^2 - u^2}},\nn\\ 
S_E &= \fr{R}{2 G}\int_{\e_{UV}}^{u_0} du \frac{u_0}{u\sqrt{f}\sqrt{u_0^2 - u^2}} ,
\label{equation}
\eeq
where a UV cutoff $\e_{UV}$ is introduced to regularize a UV divergence. 

For a pure AdS case with $f(u) =1$, the holographic entanglement entropy becomes
\al{
 &S_{AdS} = \frac{c}{3}\log\left(\frac{ l}{\epsilon_{UV}} \right) , \la{HEE:AdS}
}
where $c = 3 R/ 2G$ means a central charge of the dual CFT. This is the entanglement entropy of a two-dimensional CFT with a UV divergence. In order to discuss finite contribution, we define a renormalized entanglement entropy with removing the UV divergence
\be
S^{(re)}_E = S_E + \fr{c}{3} \log \e_{UV}  .
\ee
Then, the renormalized one is UV divergence-free. For a Schwarzschild-type AdS black hole, since the blackening factor $f(u)$ approaches one at the boundary, the leading contribution to the renormalized entanglement entropy in the UV region is given by $S_E \sim c \log l /3$ with small corrections. This logarithmic behavior universally occurs near the UV fixed point. In the IR regime, however, the renormalized entanglement entropy of a black hole shows different behavior. For a BTZ black hole geometry, the leading term of the renormalized entanglement entropy in an IR limit ($l \to \infty$) is given by \cite{Kim:2016jwu}
\al{
 &S^{(re)}_{E}  = \frac{R l}{4G u_h}  + \fr{R}{2 G} \log u_h + \fr{R}{2 G}  e^{- l /u_h}+ \cdots ,
 }
where the ellipsis indicates small quantum corrections. Recalling that $l$ corresponds to the volume of the spatially one-dimensional subsystem, we can see that the leading contribution to the IR renormalized entanglement entropy equals to the thermal entropy \eq{Result:thentropy} stored in the subsystem. Since a thermal entropy is an extensive quantity, the finite part of the IR entanglement entropy is proportional to the subsystem's volume. It was shown that this volume dependence universally appears in the black hole case. This volume dependence was called the volume law of the IR entanglement entropy \cite{Kim:2016jwu}.  This result shows that we can determine the horizon position and thermal entropy from the IR entanglement entropy. In the next sections, we further discuss how to determine the other thermodynamic quantities from the entanglement entropy data.


\section{How to reconstruct dual geometries via machine learning}

To reconstruct the dual geometry of entanglement entropy, let us  first discuss a perturbative method for later comparison with a nonperturbative construction. The perturbation approach is one of the good methods analyzing a complicated mathematical structure. However, a perturbative solution has an issue on the convergence range in which we can trust the perturbative solution \cite{Kim:2019feb,Kim:2019ewv,Kim:2019rwd,Kim:2020unz}. Due to the convergence, the perturbative method usually prohibits us from looking into a deep interior of a dual geometry. This indicates that we need a new nonperturbative method to obtain the dual geometry valid in the entire region. Despite this fact, a perturbative method is useful to see the connection between the entanglement entropy and its dual geometry at least in the UV region.

Now, we evaluate the entanglement entropy by applying  perturbative method. For an asymptotic AdS space including a Schwarzschild black hole, the metric function allows the following perturbative expansion in the asymptotic region ($u \to 0$)   
\beq
f(u) = 1+ \sum_i c_i  u^i . \label{fansatz}
\eeq
If the analytic form of a metric is known, the coefficients $c_i$ are uniquely fixed. Applying the previous holographic technique in \eqref{equation}, the subsystem size and its entanglement entropy are determined by the turning point
\beq
l(u_0)&=2u_0+\frac{\pi  c_1}{4}  u_0^2  + \left(\frac{c_1^2}{2}-\frac{2 c_2}{3}\right) u_0^3+\left( \frac{15 \pi  c_1^3}{128}-\frac{9\pi c_1 c_2}{32} +\frac{3 \pi c_3}{16}\right) u_0^4 +\cdots,
\nn\\
S_E(u_0)&=S_{AdS}(u_0)+\frac{c}{3}\left(-\frac{\pi c_1}{4} u_0 + \left(\frac{3 c_1^2}{8} -\frac{c_2}{2}\right) u_0^2 +\cdots \right),
\eeq
where $S_{AdS}$ is the entanglement entropy of a pure AdS with $c_i =0$. Combining these result, the entanglement entropy can be reexpressed as a function of the subsystem size instead of the turning point
\beq
S_E(l)=S_{AdS}(l)+\frac{c}{3}\left(-\frac{3\pi c_1}{16}  l  +\left(\frac{c_1^2}{32}+\frac{7\pi ^2 c_1^2}{512} -\frac{c_2}{24}\right)l^2 +\cdots \right). \label{persol}
\eeq
This result shows that a given geometry determines the entanglement entropy. 


When we take into account an inverse procedure, can we reconstruct the dual geometry from a given entanglement entropy? Since known entanglement entropy fixes all coefficients in \eq{persol} uniquely, it is also possible to reconstruct the dual geometry of a given entanglement entropy. However, the above reconstruction is perturbative, so that the obtained geometry is valid only in the UV  region (or small subsystems size). To go beyond the perturbation, we need to reconstruct a dual geometry nonperturbatively, This nonperturbative reconstruction is important to understand IR physics of a dual QFT and, moreover, can give us information about other physical properties. To do so, we exploit the DL technique. The DL method was also used to understand classical systems governed by position- and velocity-dependent forces \cite{Song:2020agw}.


Now, we assume that an entanglement entropy is given as the function of a subsystem size $l$, and that it follows the volume law in the large subsystem size limit. From now on, we call the given entanglement entropy a true data, $S_{true} (l)$, for convenience. In this case, since the volume law comes from the thermal entropy, we expect that the dual geometry is given by a black hole type geometry. Keeping this fact in mind, we try to reconstruct the exact dual geometry from the given entanglement entropy data.

To perform the above integrals numerically, we replace the integral range by $N$ small intervals. Here, we take $N=2000$. Then, the integrations in \eq{equation} are represented as recurrence relations between $(k-1)$-th and $k$-th layers for $k \le N$.  From now on, we follow the convention in Ref. \cite{Song:2020agw}.  Applying the fourth-order Runge-Kutta method, the recurrence relations are written as
\beq
l^{(k )} &= l^{(k-1)} +\frac{1}{6}\left(\d l(u^{(k-1)})+4 \, \d l\left(u^{(k-1)}+\frac{\Delta u}{2}\right)+\d l(u^{(k-1)}+\Delta u)\right) ,
\nn\\
S_E^{(k)} &= S_E^{(k-1)} +\frac{1}{6}\left(\d S_E(u^{(k-1)})+4 \,\d S_E\left(u^{(k-1)}+\frac{\Delta u}{2}\right)+\d S_E(u^{(k-1)}+\Delta u)\right) ,
\eeq
with
\beq
&\d l(u^{(k)})  = \fr{2u^{(k)}}{\sqrt{f(u^{(k)})} \, \sqrt{u_0^2 - ( u^{(k)})^2}} \quad  {\rm and} \quad  \d S_E(u^{(k)}) =  \fr{R}{2 G}\frac{u_0}{u^{(k)}\sqrt{f(u^{(k)})}\sqrt{u_0^2 - (u^{(k)})^2}} ,
\eeq
where $u^{(k)} =u^{(0)} + k \Delta u$ with $\D u = (u_0-u^{(0)})/N$ indicates the position of the $k$-th layer in the $u$ direction. Here $u^{(0)} =10^{-2}$ corresponds to the position of the zeroth layer which plays a role of a UV cutoff. Under this parameterization, the turning point appears at the $N$-th layer, $u_0=u^{(N)}$. When the blackening factor is given, the subsystem size and entanglement entropy are determined in terms of the turning point. After the performing the integration, in other words, the subsystem size and entanglement entropy are determined as functions of the turning point,  $l^{(N)}( u^{(N)})$ and $S_E^{(N)}( u^{(N)})$, in the holographic setup. 

In order to describe the given true data holographically, we have to find the function $f(u)$ for the dual geometry of the true data. To do so, we first identify the holographic subsystem size with that of the true data , $l^{(N)}(u^{(N)})=l$. In this case, if a testing function for $f(u)$ is really the one of the dual geometry, the holographic entanglement entropy must equal to the true data, $S_E^{(N)} (u^{(N)}) = S_{true} (l^{(N)}( u^{(N)})) $. If we choose a wrong testing function, $S_E^{(N)} (u^{(N)}) = S_{true} (l^{(N)}( u^{(N)})) $ is not satisfied. As a consequence, we can find $f(u)$ of the dual geometry by checking whether a test function satisfies $S_E^{(N)} (u^{(N)}) = S_{true} (l^{(N)}( u^{(N)}))$.

For a non-extremal black hole, the blackening factor $f(u)$ is generally factorized into $(1-u/u_h) \, g(u)$, where $g(u)$ is regular in the outside of the horizon. Using this fact, we define the following simple loss function
\be
Loss &=&  \sum_{a=1}^{M}\left | S_E^{(N)} ( u_a^{(N)}) - S_{true} (l^{(N)}(u_a^{(N)})) \right| \nn
&& + C_{reg} \, \sum_{k=1}^N \lb g (u^{(k)} )- g (u^{(k-1)} ) \rb^2 + C_{bdy} \, [g(u^{(0)} ) -1]^2 . \label{loss_eq}
\ee
Here, $u_a^{(N)}$ indicates the $a$-th turning point when we consider $M$ turning points. From now on, 
we take $M=10$ and $u_a^{(N)}= a/1.01$ with an integer $a \le M$ where the denominator $1.01$ was introduced to satisfy the constraint $u_a^{(N)} < u_h = 10$ for all $a$. This implies that we pick up ten subsystems with different sizes which are characterized by $l^{(N)}( u_a^{(N)})$. When the turning points are fixed, we can find $g(u)$ satisfying $S_E^{(N)} (u_a^{(N)}) = S_{true} (l^{(N)}( u_a^{(N)}))$ for all $a$ simultaneously by varying $g(u)$. In this case, the resulting $g(u)$ specifies the dual geometry of the true entanglement entropy data. If $M$ increases, we may obtain more accurate results. Above $C_{reg}$ and $C_{bdy}$ are two appropriate constants, which were introduced to satisfy some conditions. At the early stage, we assumed that the asymptote of the dual geometry is an AdS space, which requires $g(0) = 1$. This is automatically satisfied by minimizing the last term of the loss function. On the other hand, the second term is needed to make $g(u)$ smooth. In this work, we optimize the above loss function by applying the Adam method with $C_{reg} = 0.03$ and $C_{bdy}=1$ \cite{kingma2017adam}.

\section{Dual geometry of entanglement entropy}
Applying the DL technique discussed in the previous section, in this section we explicitly reconstruct the dual geometries when the entanglement entropy data are given. In the first two cases, the dual geometries
are known and in the last case the dual geometry is unknown.

\subsection{BTZ black hole}

\begin{figure}
\includegraphics[scale=0.27]{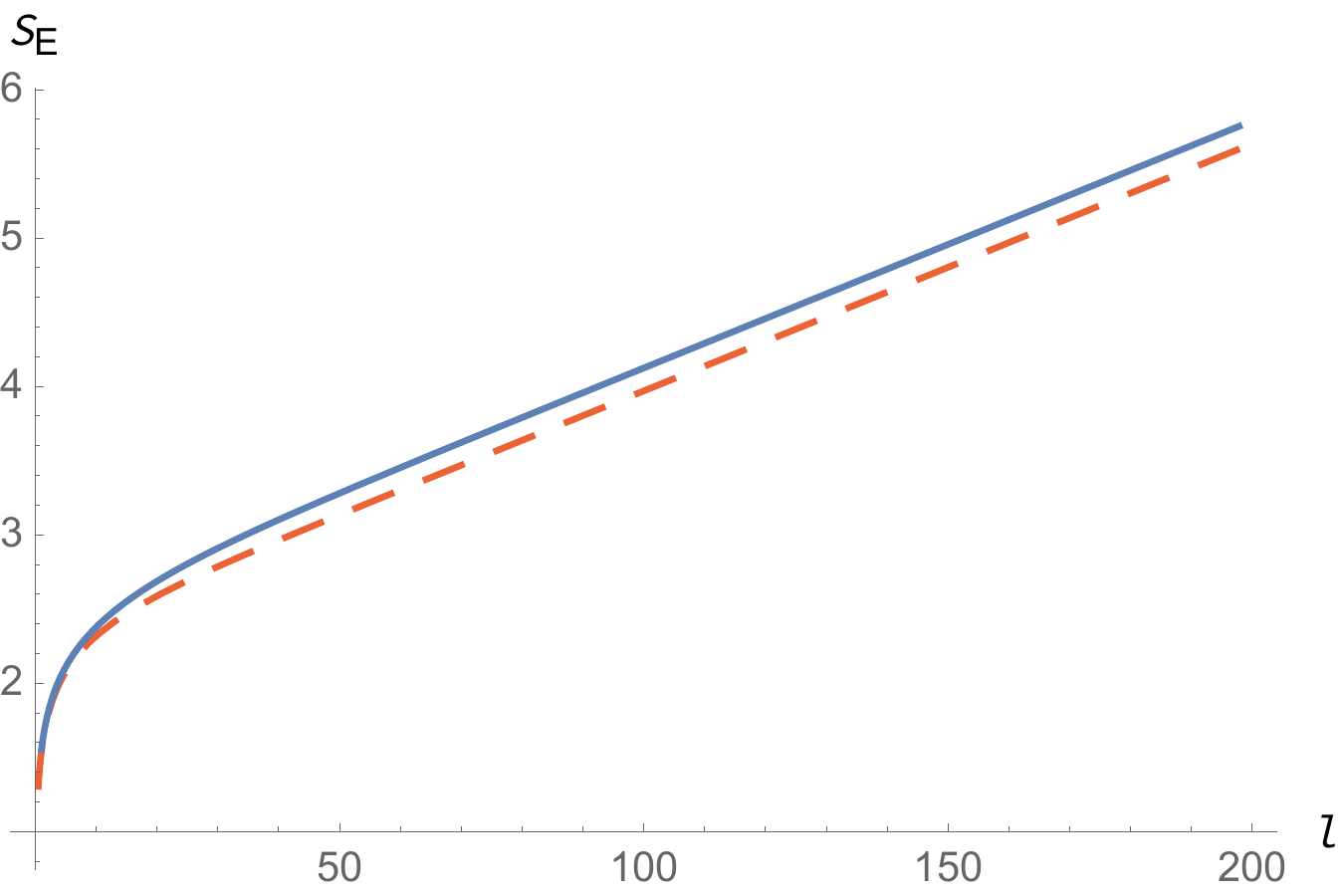}
\centering
\caption{ The entanglement entropy of two black holes,  BTZ black hole (red-dashed curve) and string cloud geometry (blue-solid curve). We take $\e_{UV}=10^{-2}$, $R=1$, $u_h=10$ and $c=1$. }
\label{data pbrane}
\end{figure}

First, we take into account the entanglement entropy of a known geometry in order to check the validity of the nonperturbative reconstruction. The BTZ black hole and its holographic entanglement entropy are analytically well-known. The blackening factor of the BTZ black hole $f(u)$ is given by
\beq
f = 1 - \left(\frac{u}{u_h}\right)^2,
\eeq
where $u_h$ is the black hole horizon. Applying the RT formula, one can easily calculate the entanglement entropy as the following form
\al{
S_{BTZ}(l) &=  \frac{c}{3}
 \log \left(\frac{2 u_h}{\epsilon_{UV}} \sinh\left(\frac{l}{2u_h}\right)\right) , \label{s bh}
}
where $\epsilon_{UV}$ means a UV cutoff. In Fig.\ref{data pbrane}, we plot the entanglement entropy of the known black hole geometries, BTZ black hole and string cloud geometry which is equivalent to the $p$-brane gas geometry for $p=1$.

\begin{figure}
\hspace{-0.5cm}
\subfigure[$g(u)$]{\includegraphics[scale=0.25]{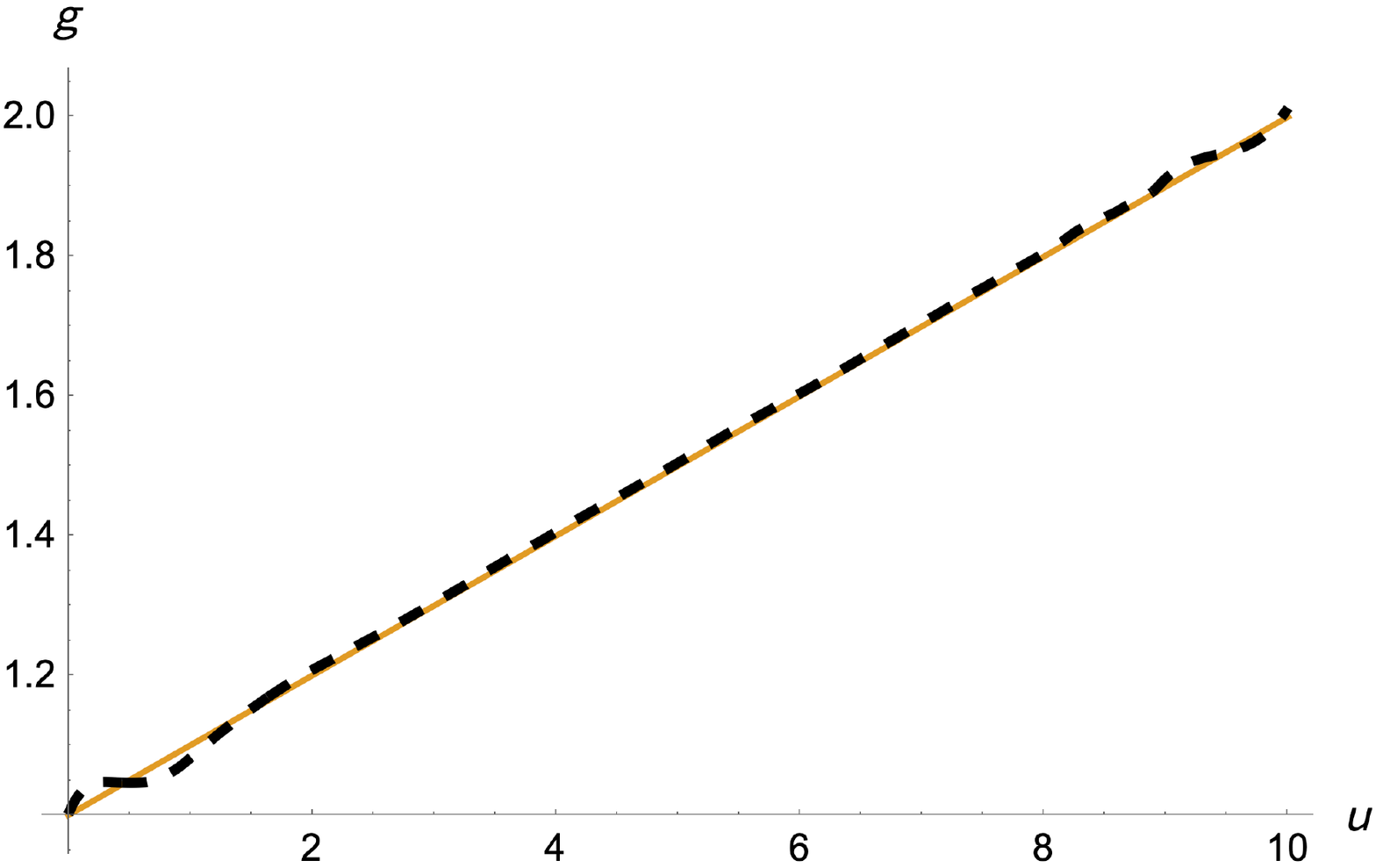}}
\hspace{0.5cm}
\subfigure[$g''(u)$]{\includegraphics[scale=0.25]{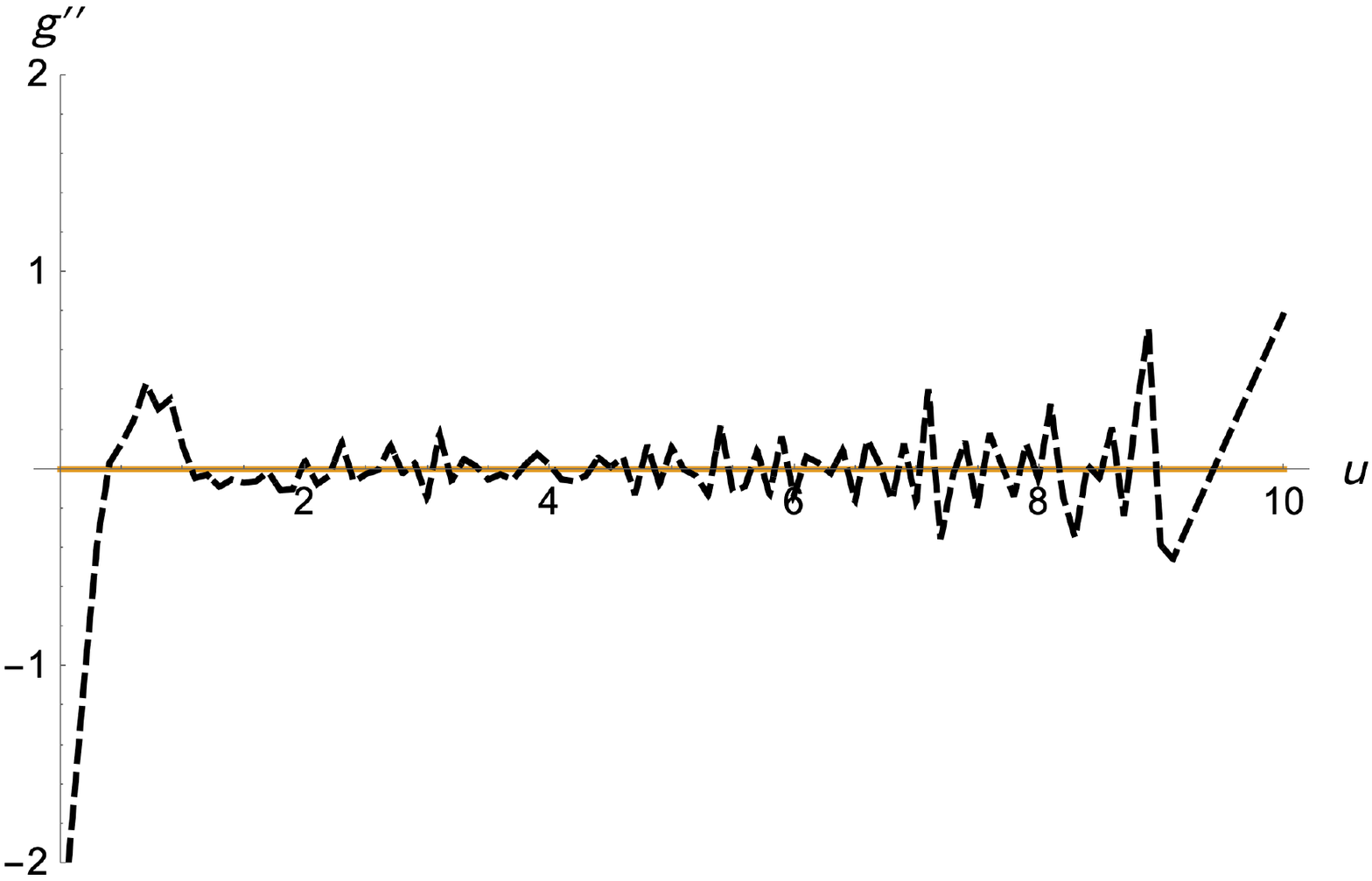}}
\centering
\caption{We plot (a) $g(u)$ and (b) $g''(u)$ (black-dashed curve) which is the numerical result obtained by the DL method. In (a), we also plot $g(u) =1 + u/u_h$ (orange-solid curve) for comparison where we take $u_h = 10$.
}
\label{bh} 
\end{figure}

Now, we assume that we have the entanglement entropy data in \eq{s bh} without knowing its dual geometry. Then can we reconstruct its dual geometry? If one can reconstruct its dual geometry, the holographic map of the obtained geometry intriguingly gives us more information about this system. In the IR region ($l \to \infty$), the entanglement entropy \eq{s bh} up to UV divergence reduces to the thermal entropy stored in the subsystem, which is proportional to the volume, $l$
\be
S_{th} (l) \approx  \frac{R}{4G} \fr{l}{u_h} .
\ee
Recalling that $l$ corresponds to the spatial volume of the dual QFT, the volume law of the IR entanglement entropy indicates that the dual geometry must be a black hole type geometry, as mentioned before. Together with the ansatz in \eq{Ansatz:SBhole}, the DL method determines the dual geometry, $g(u)$, numerically as shown in Fig.\ref{bh}. 
The result of Fig.2(a) is almost linear with some numerical error. This becomes more manifest when we calculate $g''(u)$ numerically. The resulting $g''(u)$ in Fig.2(b) is zero with small oscillating error. This indicates that $g(u)$ must be a linear function of $u$. Due to this reaon, the resulting numerical data is well fitted by the following blackening factor
\beq
f(u)=\left(1-\frac{u}{u_h}\right)\left(1.0036 + 0.9978 \frac{u}{u_h}\right). \label{ML bh}
\eeq
This DL result is consistent with the blackening factor of the BTZ black hole up to small numerical error. The numerically obtained geometry reproduces the starting entanglement entropy in \eq{s bh}.

From the numerical metric, we can determine other physical quantities of the system. For example, the obtained metric determines $g(u_h) = 2.0014$. Using this value, we see that the temperature of the system is given by
\be
T = \fr{0.1593}{u_h}  .
\ee
Moreover, we see that the system has the following internal energy densities
\beq
\r_E \equiv \fr{E}{l} &= \fr{0.0199 R }{G}\fr{1}{u_h^2}    .
\eeq
These results are consistent with the results derived from the BTZ black hole.

\subsection{String cloud geometry}

In addition to the BTZ black hole, there exists another black hole solution called the string cloud geometry \cite{Park:2021wep,Park:2021tpz,Stachel:1980zr,Chakrabortty:2011sp,Chakrabortty:2016xcb,Park:2020jio,Park:2020xho}. When open strings are uniformly distributed in an AdS space, one obtains the string cloud geometry characterized by \eq{pbrane f}. The string cloud geometry corresponds to the specific case of the previous $p$-brane gas geometry with $p=1$. The string cloud geometry has the following blackening factor
\be
f(u) = 1 - \fr{u}{u_h}  .
\ee
Computing the entanglement entropy following the RT formula, we can determine it only numerically as shown in Fig.\ref{data pbrane}, because the analytic solution is not known. Even in this case, it is still possible to reconstruct the dual geometry from the numerical data.

In the IR region of Fig.\ref{data pbrane}, the entanglement entropy of the string cloud geometry has a linear slope. This linearity indicates the volume law for the two-dimensional QFT, so that the dual geometry becomes a black hole. Recalling that the IR entanglement entropy reduces to the thermal entropy in this case, the slope in the IR region is associated with the horizon's position
\be
\fr{d S_E}{d l} = \fr{R}{4 G } \fr{1}{u_h}.
\ee
When the central charge is given by $c = 3R/2G= 1$, the slope of the IR entanglement entropy in Fig.\ref{data pbrane} determines the horizon's position to be $u_h = 10$.

\begin{figure}
\hspace{-0.5cm}
\includegraphics[scale=0.25]{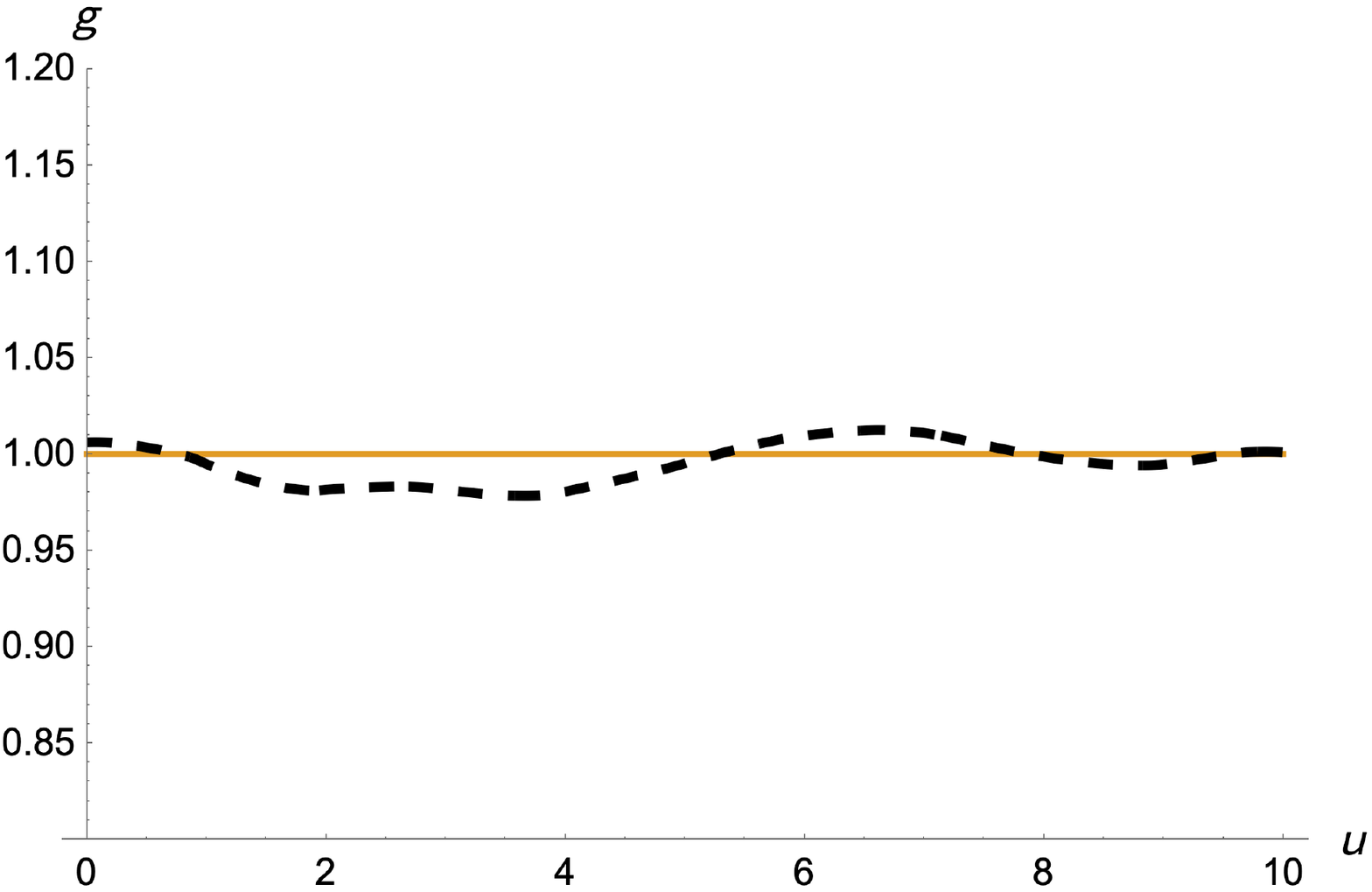}
\centering
\caption{For the string cloud geometry, we depict the numerical DL result of $g$ (black-dashed curve) where we used $u_h=10$. The result is consistent with $g=1$ up to small numerical error.
}
\label{pbrane}
\end{figure}

Applying the DL method to the entanglement entropy data (the entanglement entropy of the string cloud geometry in Fig.1), we finally determine the dual geometry numerically, as shown in Fig.\ref{pbrane} where $g(u)$ is given by a constant up to small numerical error. This geometry reproduces the entanglement entropy of the string cloud geometry. The numerical data is further well fitted by the following function
\beq
f(u) = g (u)  \left( 1 - \frac{u}{u_h}\right).
\eeq
with
\be
g (u) = 0.9953  .
\ee
This is the metric expected from the entanglement entropy and consistent with the known metric of the string cloud geometry.

Using the two quantities, $u_h$ and $g(u_h)$, obtained by the DL method, we also determine the thermodynamic quantities. The system described by the above entanglement entropy has the temperature
\be
T = 0.0079 ,
\ee
and its internal energy density is given by
\be
\r_E = \fr{0.0001 \ R}{G} .
\ee
The other quantities like free energy, pressure, and specific heat can be also determined from theses quantities by applying the thermodynamic relations studied in the previous section.

\subsection{Unknown dual geometry of entanglement entropy}
In the previous sections, we took into account the known geometries and their entanglement entropy data. Now, we look into the case having only the entanglement entropy data and reconstruct its unknown dual geometry. Let us suppose that the system has the following entanglement entropy
\beq
S_E (l)&=\frac{c}{3} \log\left(\frac{F(l) }{\epsilon_{UV}} \right)  , \la{Ansatz:QFTEE}
\eeq
where $F(l)$ is an arbitrary function satisfying two boundary conditions, $F(l) \to l$ at $l \to 0$ and $F(l) \to e^{l/a}$ with an arbitrary constant $a$ at $l \to \infty$. Here, the first condition requires the existence of a UV fixed point. On the other hand, the second condition was imposed to obtain the volume law in the IR limit. Theses two boundary conditions restrict the dual geometry to a black hole. 

Now, we take into account the following simple example
\beq
S_E (l)&=\frac{c}{3}  \log\left( \frac{a}{\epsilon_{UV}}\left(\exp\left(\frac{l}{a}\right) -1\right)\right),\label{s exp} 
\eeq
which satisfies the required boundary conditions. If we ignore the UV divergence part, the leading term of the IR entanglement entropy is given by
\be
S_E \approx \fr{c}{3} \fr{l}{a} .
\ee
Here, the volume law of this IR entanglement entropy is caused by the thermal entropy. Recalling the following relation $c=3 R /(2 G)$, we see that the horizon in the dual geometry appears at $u_h = a/2$.

Applying the perturbative method, the perturbative expansion of the entanglement entropy in the UV region determines the metric as the following series
\beq
f(u) = 1-\frac{4 }{3 \pi  }\frac{u}{u_h}+\left(\frac{1}{3}+\frac{4}{3 \pi ^2}\right) \frac{u^2}{u_h^2}-\left(\frac{146}{567 \pi }+\frac{32}{27 \pi ^3}\right)\frac{u^3}{u_h^3}+\cdots
\nn\\
=\left( 1- \frac{u}{u_h}\right) \left(1+\frac{0.575587 u}{u_h}+\frac{1.04402 u^2}{u_h^2}+ \frac{0.923828 u^3}{u_h^3}+\cdots\right).
\eeq
This perturbative result is valid only in the UV region ($u/u_h \ll 1$), so that it does not give us information about the IR physics. Due to this reason, the perturbative calculation cannot determine the black hoe geometry correctly. To overcome this problem and to know IR physics, we have to exploit a nonperturbative method.

\begin{figure}
\hspace{-0.5cm}
\subfigure[$g$]{\includegraphics[scale=0.28]{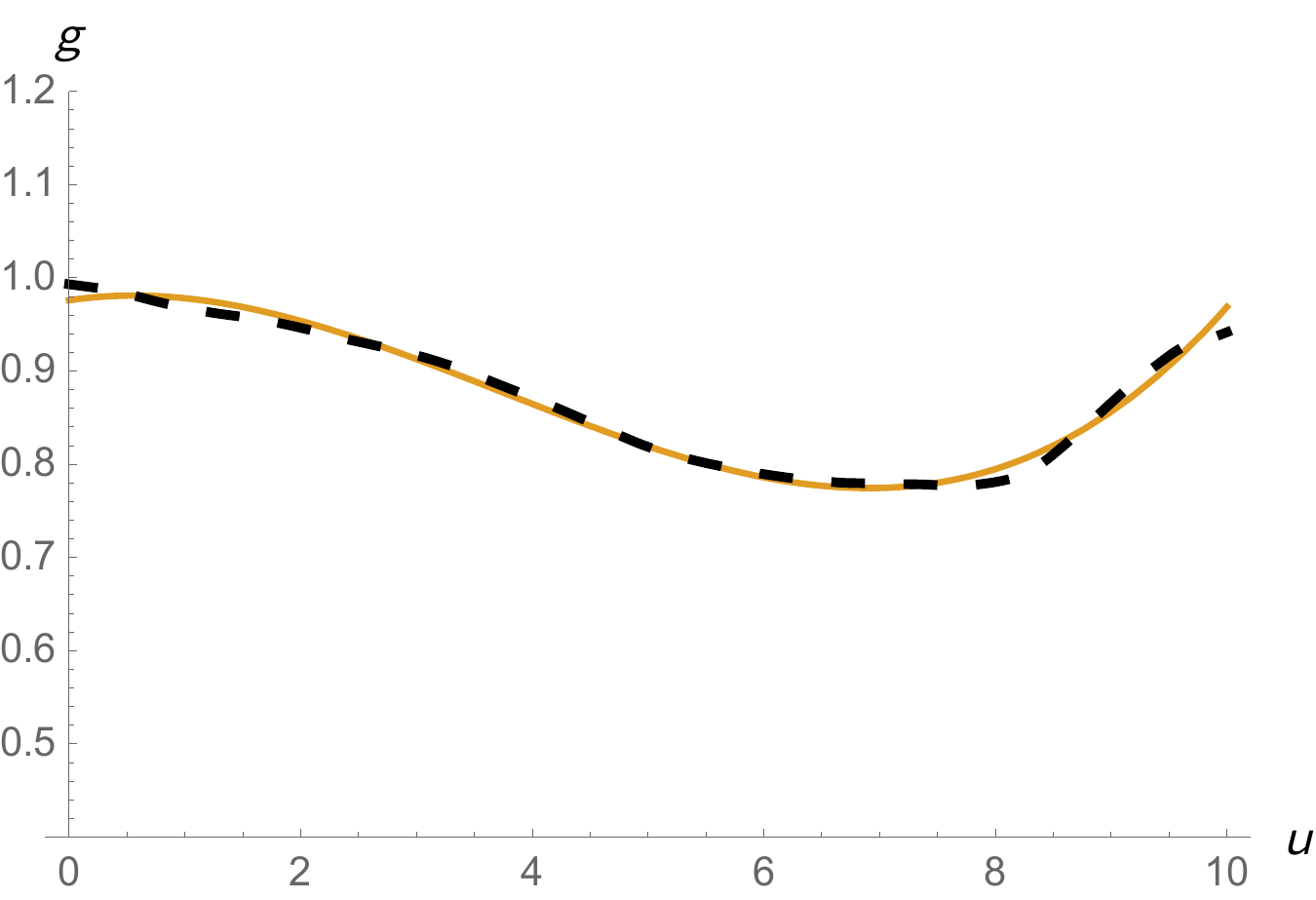}}
\subfigure[$dg/du$]{\includegraphics[scale=0.28]{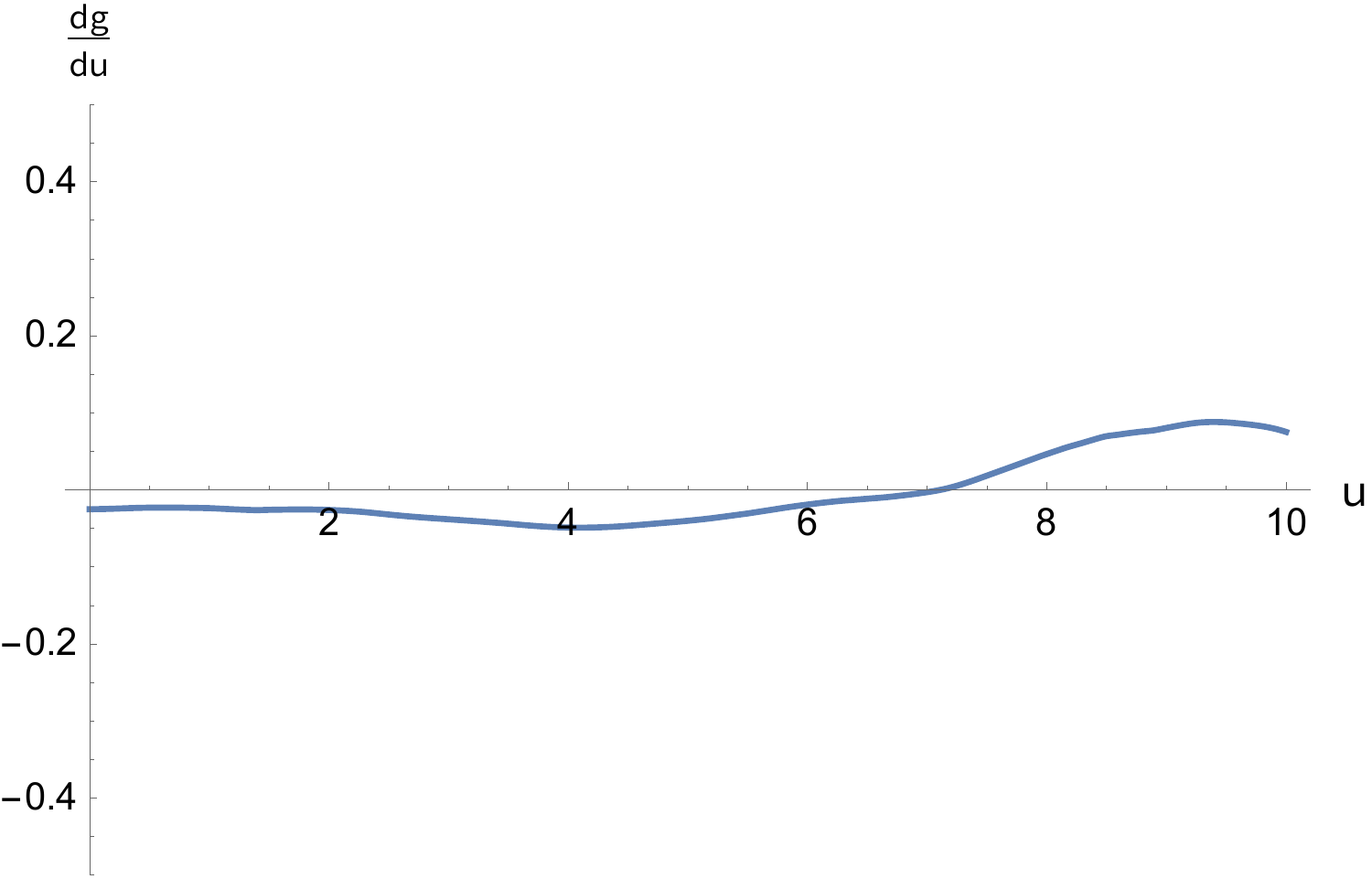}}
\centering
\caption{We plot the numerical DL results of (a) the regular part of the blackening factor $g$ (black-dashed curve) and (b) its derivative $dg/du$. They are derived from the entanglement entropy data whose dual gravity is not known. The orange curve in (a) indicates a nonperturbative approximation fitting the DL result up to $u^4$ order. 
}
\label{fig exp}
\end{figure}

Applying the nonperturbative DL technique, we obtain the following dual geometry
\beq
f(u) = \left(1-\frac{u}{u_h}\right) \ g(u) , \label{ML cosh} %
\eeq
with a numerical function $g(u)$ in Fig.\ref{fig exp}(a). The value of $g(u)$ at the horizon is given by $g(u_h) = 0.471$ for $a=20$. This value together with the horizon determines the temperature and internal energy of the considered system 
\beq
T & = \frac{0.471}{2 \pi} \frac{1}{u_h} ,
\nn\\
E & = \frac{0.471 \ c}{24 \pi }\frac{l}{u_h^2}  .  \la{resuts:Nthermalq}
\eeq
These results show that the internal energy is proportional to the degrees of freedom, as expected and that the system considered here follows the Stefan-Boltzmann law.

\begin{figure}
\hspace{-0.5cm}
\subfigure[$f$]{\includegraphics[scale=0.28]{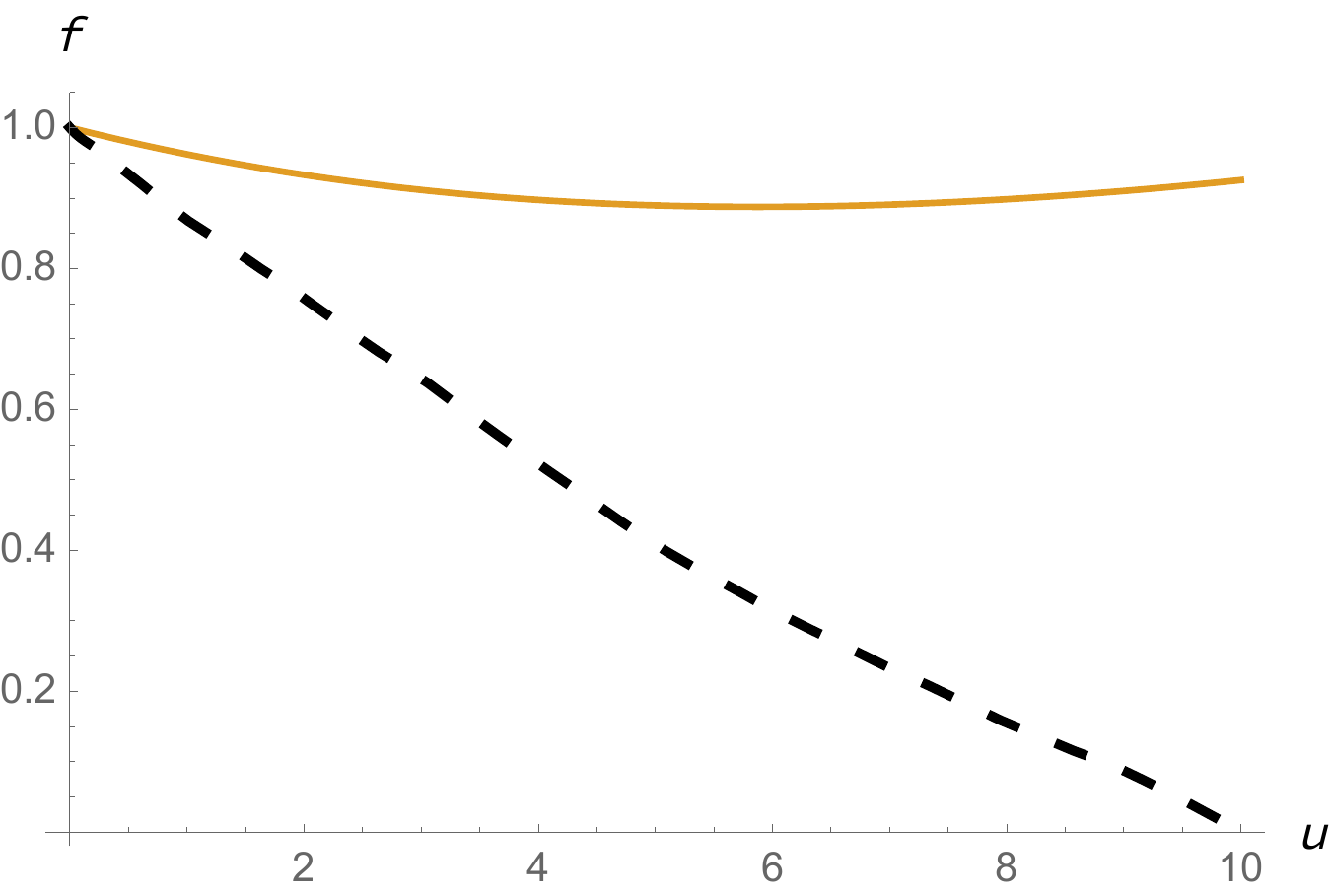}}
\subfigure[$S_E$]{\includegraphics[scale=0.28]{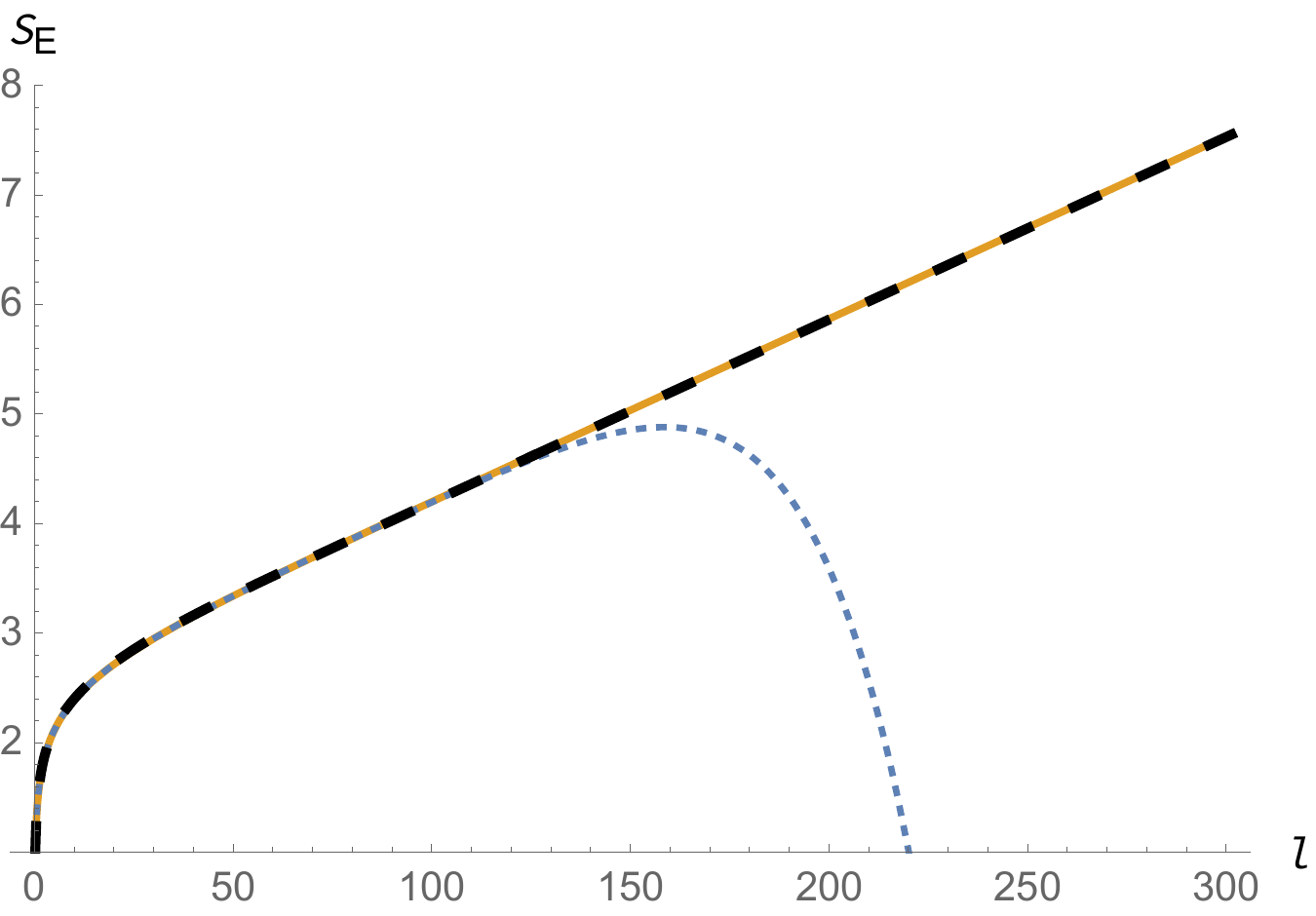}}
\centering
\caption{ (a) We plot the blackening factors evaluated by the perturbation method (orange-solid curve) and the DL method (black-dashed curve). (b) We depict the holographic entanglement entropies derived from the original one (orange-solid curve), DL result (black-dashed curve), and perturbative result (blue-dotted curve). The DL result reproduces the original entanglement entropy, while the perturbative result is valid only in the small $l$ region, as mentioned before.  
}
\label{exp}
\end{figure}

From the numerical metric, intriguingly, it is also possible to find a nonperturbative approximation valid in the outside of the black hole. For example, the numerical data, as shown in Fig.\ref{fig exp}(a), is well fitted by the following polynomial 
\be
g(u) = 0.975+0.186\frac{ u}{u_h}-1.805\frac{ u^2}{u_h^2}+1.611\frac{ u^3}{u_h^3}  .
\ee
This analytic function reproduces the starting entanglement entropy \eq{s exp} up to a small numerical error. The numerical and analytic results give rise to the almost the same metric and entanglement entropy, as shown in Fig.\ref{exp}. Moreover, the thermodynamic quantities derived from the analytic function leads to the almost same result as \eq{resuts:Nthermalq}
\beq
T  & = \frac{0.484}{2\pi} \fr{1}{u_h},
\nn\\
E  & = \frac{0.484 \ c}{24 \pi } \fr{l}{u_h^2}.
\eeq
These results intriguingly show that the dual geometry reconstructed from the entanglement entropy data gives us more physical information on the considered system.

\section{Discussion}

We studied how to reconstruct the dual geometry of entanglement entropy data via the deep learning method. After making a neural network structure of the Ryu-Takayanagi formula, we find the dual geometry reproducing given entanglement entropy data. In this work, we focused on specific entanglement entropy which is linearly proportional to the subsystem's volume in the large size limit. This IR feature generally occurs when the entanglement entropy flows to a thermal entropy in the IR region. In this case, the dual geometry must be a black hole-type geometry. By applying the deep learning method studied here, we reconstructed the known black hole solutions, BTZ black hole and string cloud geometry, from the analytic and numerical entanglement entropy data. We also took into account arbitrary entanglement entropy whose holographic dual is not known. Even in this case, we successfully reconstructed the dual geometry which reproduces the starting entropy data. 

Reconstructing the dual geometry from entanglement entropy data is important to understand other physical properties of the same system. Since the dual geometry can provide more information about the underlying structure of the dual QFT, it allows us to figure out other physical quantities beyond reproducing the original entanglement entropy. From the dual geometry of the entanglement entropy, we extracted information about thermodynamic variables like temperature and internal energy which characterize thermal properties of the system in the IR limit.
 
In the present work, we concentrated on black hole geometries because the entanglement entropies of their dual QFT's have an universal feature in the IR region. However, the entanglement entropy RG flow of a general QFT does not always admit thermodynamics in the IR region. In this case, can we reconstruct its dual geometry from the entanglement entropy data? In general, a nontrivial RG flow of the entanglement entropy is crucially related to the change of couplings. Therefore, if we know the entanglement entropy as well as the $\b$-functions of system's couplings, these RG data may enable us to reconstruct the dual geometry beyond the black hole geometries studied here. We hope to report more results on this issue in future works.

\vspace{1cm}

{\bf Acknowledgement}

We would like to thank N. Kim, H. Kim and J. Lee for valuable discussion and comments. CP was supported by Mid-career Researcher Program through the National Research Foundation of Korea grant No. NRF-2019R1A2C1006639. SJK was supported by Basic Science Research Program through the National Research Foundation of Korea, funded by the Ministry of Education grant No. NRF-2021R1I1A1A01052821.


\bibliographystyle{apsrev4-1.bst}
\bibliography{ref.bib}

\end{document}